\documentclass[letterpaper]{JHEP}
\usepackage{epsf}

\def\d{\partial}
\def\db{\bar\partial}

\newcommand{\be}{\begin{eqnarray}}
\newcommand{\ee}{\end{eqnarray}}
\newcommand{\beq}{\begin{equation}}
\newcommand{\equ}{\end{equation}}
\newcommand{\eeq}[1]{\label{#1}\end{equation}}
\newcommand{\ber}{\begin{eqnarray}}
\newcommand{\eer}[1]{\label{#1}\end{eqnarray}}

\newcommand{\abs}[1]{\left| #1 \right|}
\newcommand{\bra}[1]{\left\langle{#1}\right|}
\newcommand{\ket}[1]{\left|{#1}\right\rangle}

\newcommand{\proj}[1]{\left| #1\right\rangle\left\langle #1\right|}
\newcommand{\tr}{\rm{Tr}}

\newcommand{\ie}{{\it i.e.}}

\renewcommand{\a}{{\alpha}}
\renewcommand{\b}{{\beta}}
\renewcommand{\c}{{\gamma}}
\newcommand{\de}{{\delta}}
\newcommand{\De}{{\Delta}}
\renewcommand{\l}{{\lambda}}
\renewcommand{\k}{{\kappa}}
\renewcommand{\th}{{\theta}}

\preprint{hep-th/0104017\\
YITP-SB-01-13}

\author{Leszek Hadasz${}^{1,2}$,~Ulf Lindstr\"{o}m${}^{3}$,~
Martin Ro\v{c}ek${}^{1}$,~ and Rikard von Unge${}^{4}$\\ ~ \\
${}^{1}$C.N. Yang Institute for Theoretical Physics,
State University of New York\\
Stony Brook, NY 11794-3840, USA\\ ~ \\
${}^{2}$M. Smoluchowski Institute of Physics, Jagiellonian University\\
Reymonta 4, 30-059 Cracow, Poland\\ ~ \\
${}^{3}$Institute of Theoretical Physics, University of Stockholm\\
Box 6730\\
S-113 85 Stockholm, Sweden\\ ~ \\
${}^{4}$Institute for Theoretical Physics and Astrophysics\\
Faculty of Science, Masaryk University\\
Kotl\'{a}\v{r}sk\'{a} 2, CZ-611 37, Brno, Czech Republic\\ ~ \\
\email{leszek@insti.physics.sunysb.edu}\\
\email{ul@physto.se}\\
\email{rocek@insti.physics.sunysb.edu}\\
\email{unge@physics.muni.cz}}
\abstract{We find the
$N$-soliton solution at infinite $\th$, as well as the metric on the
moduli space corresponding to spatial displacements of the solitons. We
use a perturbative expansion to incorporate the leading
$\th^{-1}$ corrections, and find an effective short
range attraction between solitons. We study the stability of various solutions.
We discuss the finite $\th$ corrections to scattering, and find metastable
orbits. Upon quantization of the two-soliton moduli space, for any finite
$\th$, we find an $s$-wave bound state.}

\title{Noncommutative Multisolitons: \\ {\large Moduli Spaces,
Quantization, Finite $\th$ Effects and Stability}}

\begin{document}
\section{Introduction}
Recently \cite{basic} it was realized that one can construct stable
soliton solutions in noncommutative scalar field theory even though
such solitons do not exist in commutative scalar theories in higher than two
dimensions. The solutions are particularly simple when the noncommutativity
parameter $\th\to\infty$, where one finds an infinite dimensional moduli space
of solitons. This program has also been extended to noncommutative gauge
theories \cite{gaugesolI,gaugesolII,gaugesolIII,gaugesolIV,gaugesolV}.

These solitons have found an application in the context of tachyon
condensation where D-branes can be found as soliton solutions on
higher dimensional non-BPS D-branes. By turning on a B-field one makes
these non-BPS D-branes noncommutative and the soliton configurations
studied represent various types of lower dimensional D-branes
\cite{bransolI,bransolII}.

There also seems to be a place for application of these solitons in a
noncommutative description of the Quantum Hall Effect
\cite{QHEI,QHEII,QHEIII,QHEIV,QHEV,QHEVI,QHEVII}.

Motivated by these developements, we have studied what happens when 
one scatters
noncommutative solitons.  In \cite{us} we analyzed this questions using moduli
space techniques, and found a  K\"{a}hler metric on the moduli space
somewhat analogous to the metric on the moduli space of two magnetic monopoles.
A natural generalization of the results in \cite{us} is to find the moduli
space metric for $N$ solitons. In this paper we find a simple and elegant
expression for the K\"ahler potential for the general case.

The analysis in \cite{basic} was mainly done at infinite
$\th$ but a program to find corrections to the solitons at finite
$\th$ was initiated. This was followed by studies at finite $\th$, both
numerically \cite{numI,numII} and theoretically \cite{stabI,stabII}. This topic
is important, since one would like to know if the solitons are
stable at finite $\th$, and if they are, how many of the infinite number of
moduli survive. We study this issue and find that at finite $\th$, nonradial
excitations, which were ignored in \cite{numI,numII,stabI,stabII},
destabilize all ``excited'' soliton states, and leave only the basic
$N$-multisoliton solutions.

Some quantum issues have also been studied in \cite{quantI,quantII}. However,
the discussion in \cite{quantI} involves averaging over nonradial modes, which
we find play an essential role.

This paper is organized as
follows: In section 2, we construct the general
$N$-soliton solution at infinite $\th$. We find the metric on the moduli space
corresponding to spatial displacements of the solitons, and discuss the
three-soliton case in detail. In section 3, we introduce a perturbative
expansion that allows us to incorporate the leading $\th^{-1}$ corrections. In
the two-soliton case we find an effective short range attraction between
solitons. In section 4 we use these perturbative results to study the
stability of various solutions. In section 5 we focus on the two-soliton case
and discuss the finite $\th$ corrections to scattering. We find a range of
interesting phenomena, including metastable orbits. In section 6, we quantize
on the two-soliton moduli space, and find an $s$-wave bound state for any
finite $\th$.

While writing of this paper, we became aware that results which
have some overlap with our results were presented in \cite{talk}.

\section{Multisolitons at infinite $\th$}
\subsection{Multisoliton solutions}
The two-soliton solution at infinite $\th$, constructed in
\cite{basic}, is
\beq
\label{two_sol_0}
\Phi_2=\l\left(\proj{z_+} + \proj{ z_-}\right)~,
\equ
where $\l$ is an extremum of the potential $V(\phi),$
\beq
\ket{z_{\pm}} =
\frac{\ket{z} \pm \ket{-z}}{\sqrt{2\left(1\pm{\rm e}^{-2|z|^2}\right)}},
\equ
and $\ket{\pm z} = {\rm e}^{-\frac12|z|^2 \pm za^{\dag}}\ket0$. This can
be generalized to the $n$ soliton case as follows: Let $z_\a, \a=
1,\ldots,n$ be pairwise different complex numbers satisfying the center
of mass condition $\sum\limits_{\a=1}^{n}z_\a = 0.$ With
\beq
\ket{z_\a}  \equiv
{\rm e}^{-\frac12|z_\a|^2 + a^\dagger z_\a}\ket0~,
\equ
the multi-soliton solution is
\beq
\label{mult_sol_0}
\Phi_n=\l\sum\limits_{\a ,\b =1}^{n}
\left|z_\a\right\rangle A^{-1}_{\a \b}
\left\langle z_\b\right|\equiv \l P_n~,
\equ
where $A$ is the $n\times n$ matrix,
\beq
A_{\a\b} = \left\langle z_\a\left|\right.z_\b\right\rangle =
e^{-\frac12(|z_\a - z_\b |^2+z_\a\bar z_\b-z_\b\bar z_\a)}~,
\equ
and $P_n$ is a rank $n$ projection operator onto the linear subspace
of the harmonic oscillator Hilbert space ${\cal H}$
spanned by the vectors $\ket{z_\a}$.

For large separations, \ie,  $\abs{z_\a-z_\b}\gg 1$ for all
$\a\neq\b$,
\beq
\Phi_n = \sum_{\a = 1}^n \l\ \proj{z_\a} +{\cal O}\left({\rm e}^{-
\abs{z_\a-z_\b}^2}\right)~,
\equ
\ie, in this limit, $\Phi_n$ describes $n$ well separated level 0
solitons. To study the limit $z_\a\to 0$ it is convenient to
introduce a new basis:
\begin{eqnarray*}
\ket{u_1} & = & {\rm e}^{\frac12\epsilon^2|z_1|^2}\ket{\epsilon z_1} =
		\ket{0} + {\cal O}(\epsilon)~, \\
\ket{u_2} & = & \frac{1}{\epsilon(z_2-z_1)}\left(
	{\rm e}^{\frac12\epsilon^2|z_2|^2}\ket{\epsilon z_2}-
\ket{u_1}\right) =
      \ket{1} + {\cal O}(\epsilon)~,\\
\ket{u_3} & = & \frac{\sqrt{2}}{\epsilon^2(z_3-z_1)(z_3-z_2)}\left(
	{\rm e}^{\frac12\epsilon^2|z_3|^2}\ket{\epsilon z_3}- \ket{u_1} -
		\epsilon(z_3-z_1)\ket{u_2}\right) =
		\ket{2} + {\cal O}(\epsilon)~, \\
& \vdots &
\end{eqnarray*}
where $\epsilon$ is a small parameter and $\left|i\right\rangle =
\frac{1}{\sqrt i!}\left(a^{\dag}\right)^i\left|0\right\rangle.$
$P_n$ can now be written
as
$$
P_n = \sum_{\a,\b = 1}^n
\ket{u_\a}\bra{u_\a}u_\b\rangle^{-1}\bra{u_\b}
      = \sum_{i=0}^{n-1}\ket{i}\bra{i} + {\cal O}(\epsilon)~,
$$
and hence, in the limit $\epsilon \to 0$,
$\Phi_n$ describes $n$ solitons from the $0$ up to the $n-1$ harmonic
oscillator level, all at the origin.

In the generic case of different $z_\a,$  $P_n$ is unitarily equivalent
to the projector onto the subspace spanned by the
vectors $\left|i\right\rangle$  with $i < n$. To
construct the unitary transformation explicitly, we diagonalize the matrix
$A$ (which is hermitean and positive semidefinite). Let $\vec v_{(\a)}$ and
$a_{(\b)}$ denote its (orthonormalized) eigenvectors and corresponding
eigenvalues:
\beq
A_{\a\b}\,v_{\b(\c)}=a_{(\c)}\,v_{\a(\c)}~,\qquad \vec v_{(\a)}\cdot\vec
v_{(\b)} = \de_{\a\b}~.
\eeq{diaga}
Defining
\beq
W_{\a\b} = \frac{v_{\a(\b)}}{\sqrt{a_{(\b)}}}
\equ
and the Hilbert space vectors
\beq
\left|w_\a\right\rangle = \sum_{\b=1}^{n}\left|z_\b\right\rangle
W_{\b\a}~,
\equ
we find
$$
\left\langle w_\a|w_\b\right\rangle = \de_{\a\b}~,
\hskip 1cm
{\rm and}
\hskip 1cm
\Phi_n = \l\sum_{\a = 1}^n\left|w_\a\right\rangle
\left\langle w_\a\right|~.
$$
For $n=2$ this procedure -- as expected -- gives
$\{\left|w_1\right\rangle,\left|w_2\right\rangle\} =
\{\left|z_+\right\rangle,\left|z_-\right\rangle\}$.

Finally, we take any orthonormal basis in ${\cal H}$ whose first $n$
vectors coincide with $\left|w_\a\right\rangle$ and denote it by
$\{\left|w_j\right\rangle\}.$ For
\beq
U = \sum_{j=1}^\infty\left|w_j\right\rangle\left\langle j-1\right|
\equ
we have
$$
UU^{\dag} = U^{\dag}U = \mbox{\boldmath $1$}~,
$$
and
\beq
\Phi_n = \l U\left(
\sum_{i=0}^{n-1}\left| i\right\rangle\left\langle i\right|\right)U^{\dag}.
\equ

In a completely analogous manner, one may construct excited multisoliton
solutions from states $\ket{z_\a,n}=(a^\dag)^n\ket{z_\a}$ (the excited
two-soliton case was worked out in detail in \cite{us}). However, as
shown in section 4, such solitons are all unstable for any nonzero
$\th$, and hence we do not discuss them further.

\subsection{Moduli spaces}
The metric on the multi-soliton moduli space is K\"{a}hler for
any $n.$ Up to a constant normalization factor we have \cite{us}:
\be
g_{z_\a z_\b} & = & \frac{1}{\l^2}{\rm Tr}
\left(\d_{z_\a}\Phi_n\d_{z_\b}\Phi_n\right)~, \nonumber \\
g_{z_\a \bar z_\b} & = & \frac{1}{\l^2}{\rm Tr}
\left(\d_{z_\a}\Phi_n\d_{\bar z_\b}\Phi_n\right)~, \\
g_{\bar z_\a \bar z_\b} & = & \frac{1}{\l^2}{\rm Tr}
\left(\d_{\bar z_\a}\Phi_n\d_{\bar z_\b}\Phi_n\right)~.
\nonumber
\ee
Straightforward calculation gives
$$
g_{z_\a z_\b} = g_{\bar z_\a \bar z_\b} = 0~,
$$
while
\beq
\label{nsol_metric}
g_{z_\a \bar z_\b} =
A^{-1}_{\a\b}\left(A_{\b\a} + \bar z_\b A_{\b\a} z_\a
- \sum_{\c,\de = 1}^n A_{\b\c}z_\c A^{-1}_{\c\de}\bar z_\de
A_{\de\a}\right) =
\d_{z_\a}\d_{\bar z_\b}
K\left(\mbox{\boldmath $z$},\bar{\mbox{\boldmath $z$}}\right)
\equ
with
\be
K\left(\mbox{\boldmath $z$},\bar{\mbox{\boldmath $z$}}\right) =
\sum_{\a = 1}^n\left|z_\a\right|^2 + \log\det A~.
\ee

\subsection{The three-soliton case}
As a simple example, we consider the three-soliton metric in detail; the matrix
$A$ becomes
\be
       \left(\begin{array}{ccc}
       1 & e^{-\frac{1}{2}\abs{z_1}^2 - \frac{1}{2}\abs{z_2}^2+\bar{z}_1 z_2} &
       e^{-\frac{1}{2}\abs{z_1}^2 - \frac{1}{2}\abs{z_3}^2+\bar{z}_1 z_3} \\
       {\rm c.c.} & 1 &
       e^{-\frac{1}{2}\abs{z_2}^2 - \frac{1}{2}\abs{z_3}^2+\bar{z}_2 z_3} \\
       {\rm c.c.} & {\rm c.c.} & 1
\end{array}\right)~,
\ee
and gives the K\"{a}hler potential (using that $z_3 = -z_1
-z_2$)
\ber
       K &=& \ln\left(
e^{2\abs{z_1}^2+2\abs{z_2}^2+\bar{z}_1 z_2 + \bar{z}_2 z_1} +
e^{-\abs{z_1}^2-\abs{z_2}^2+\bar{z}_1 z_2 -2 \bar{z}_2 z_1} +
e^{-\abs{z_1}^2-\abs{z_2}^2-2\bar{z}_1 z_2 + \bar{z}_2 z_1}
\right. \nonumber\\ &&\left.
\ \ \ \ -\, e^{-2\abs{z_1}^2 + \abs{z_2}^2 - \bar{z}_1 z_2 - \bar{z}_2 z_1} -
e^{\abs{z_1}^2 - 2\abs{z_2}^2 - \bar{z}_1 z_2 - \bar{z}_2 z_1} -
e^{\abs{z_1}^2+\abs{z_2}^2 +2 \bar{z}_1 z_2 +2 \bar{z}_2 z_1}
\right)~.\nonumber\\
\eer{3kpot}
This has the two-soliton metric as a subspace: If we take one
of the solitons (say $z_2$) far away and fix its position we can
choose the coordinates to be
\be
       z_1 &=& -\frac{z_2}2 + \zeta \nonumber\\
       z_3 &=& -\frac{z_2}2 - \zeta~,
\ee
where $\zeta$ is the
relative distance between the soliton at $z_1$ and the soliton at
$z_3$ and is taken to be much smaller than $z_2$. Inserting this in
(\ref{3kpot}) we get
\be
       K \to \frac32\abs{z_2}^2+\ln\left(2\sinh(2\bar{\zeta}\zeta)\right) ~.
\ee
Thus the geometry factorizes into two pieces, one, coordinatized by $z_2$,
which is flat, and another, coordinatized by $\zeta$, with precisely the
K\"{a}hler potential of the 2 soliton moduli space, including the conical
singularity when the solitons coincide.

We can also study what happens when all three solitons come
together. We study the most symmetrical case where the solitons are at
the same distance from the origin and separated by the angle
$\frac{2\pi}{3}$. In that case we can choose the coordinates
\be
       z_1 &=& \zeta, \nonumber\\
       z_2 &=& \omega \zeta, \\
       z_3 &=& \omega^2 \zeta, \nonumber\\
       \omega &=& e^{\frac{2\pi i}{3}}~. \nonumber
\ee
For small values of $\zeta$ we get the K\"{a}hler potential
\be
       K = \ln\left(\abs{\zeta}^6 + \frac{\abs{\zeta}^{12}}{120}
+\ldots\right)~,
\ee
giving rise to a conical singularity of the type
\be
       ds^2 = r^4\left(dr^2 + r^2 d\th^2\right)~.
\ee
This lead to a scattering angle of $\frac{2\pi}{3}$.

\section{Finite $\th$: perturbation Theory}
\label{Perturbation_Theory}
We now consider the finite $\th$ corrections to the
soliton solutions.

At finite $\th$ one has to include the derivative terms in the
energy functional:
\ber
       E = \frac{2\pi}{g^2}\tr\left(
\frac{1}{2}[a\, ,\phi ] [\phi\, ,a^{\dagger}\, ] +
\th V\left(\phi\right)\right).
\eer{energy}
If we make the ansatz
\be
\phi = \l\left(P+\frac{1}{\th} B\right)
\ee
and use the conditions $V'(0) = V'(\l) = 0$, then (\ref{energy})
gives\footnote{No order $\th^{-2}$ terms in $\phi$ contribute
to the energy at order $\th^{-1}$ because they are multiplied by $V'(\l)$ or
$V'(0)$, which both vanish.}
\be
\label{en_gen}
&&  E = \frac{2\pi}{g^2}\tr\Bigg[ \th V(\l)P +
       \frac{\l^2}{2}[\,a\, ,P\, ][P,a^{\dagger}\,] +
       \frac{\l^2}{\th}\Bigg(B\, [[P,a^\dagger\, ],a\, ]
      \\
&&
\hskip 15mm
      + \;
\frac12V''(\l)PBPBP
+ \frac12V''(0)
(1-P)B(1-P)B(1-P)\Bigg)
\Bigg] + {\cal O}\left(\th^{-2}\right)~.\nonumber
\ee
Extremizing with respect to the perturbation
$B$, to leading order we find
\be
\label{eom_pert}
V''(\l) PBP +
V''(0)\left(1-P\right)B\left(1-P\right) &=&
[\, a\, ,[\, P,a^{\dagger}\, ]] \equiv \De_P~.
\ee
This implies that $P$ has to satisfy the consistency condition
\be
\label{cons_cond}
[P,\De_P\, ]=0~.
\ee
Fortunately, this relation is fulfilled for all the soliton
solutions we consider.

We can write the solution of (\ref{eom_pert}) in the form
\be
\label{pert_sol_1}
B = \frac{1}{V''(\l)} P\De_P P +\frac{1}{V''(0)}(1-P)\De_P(1-P)
+ PX(1-P) + (1-P)X^{\dag}P~,
\ee
where $X$ is an arbitrary operator; it drops out of both
(\ref{eom_pert}) and -- as a consequnece of (\ref{cons_cond}) --
the energy (\ref{en_gen}) to order $\th^{-1}$, and hence
we choose $X=0$ in what follows.

A special class of projectors that satisfies (\ref{cons_cond}) is given
by solutions of the equation
\be
\label{cons_strong}
(1-P)aP = 0~.
\ee
It includes the projectors that give the $n$
soliton solutions (\ref{mult_sol_0}).

It is trivial to see that (\ref{cons_strong}) implies (\ref{cons_cond}),
while the reverse implication does not hold -- the projector
$|n\rangle\langle n|$ with $n > 0,$ satisfies (\ref{cons_cond}) but
not (\ref{cons_strong}).

For operators $P$ satisfying (\ref{cons_strong}), the energy
(\ref{en_gen}) can be rewritten in a simpler form
\be
\label{en_sim}
       E & = & \frac{2\pi}{g^2}\tr\left( \th V(\l)P +
       \frac{\l^2}{2}P
\right) \\
&&
- \;\frac{\pi\l^2}{\th g^2}\left(\frac{1}{V''(\l)}
+ \frac{1}{V''(0)}\right){\rm Tr}\left[
2(Pa^\dag a)^2  -2P(a^\dag\,)^2a^2 + P\right]~.
\nonumber
\ee

\subsection{Explicit solutions}
For one-soliton states, our perturbative expansion reproduces the results
in Appendix A of \cite{basic} to the appropriate order in $\th^{-1}$.

For $P = \ket{z+}\bra{z+} + \ket{z-}\bra{z-}$
(\ref{two_sol_0}), the simplifying condition (\ref{cons_strong}) is
satisfied. One may calculate the first order correction $B$ to $\Phi$
from (\ref{pert_sol_1}):
\be
\label{B_P}
B = \frac{1}{V''(\l)} P\De_P P +\frac{1}{V''(0)}(1-P)\De_P(1-P)~,
\ee
by substituting $\De_P=[\, a\, ,[\, P,a^{\dagger}\, ]]$ as
above, but to find the corrections to the energy, it is easier to
use the expression (\ref{B_P}), simplify using the cyclicity of the
trace, and the condition (\ref{cons_strong}), and only then substitute
the explicit form of $P$.

At order $\th$ the energy is a constant:
\be
E_0 = \frac{2\pi}{g^2}2\th V(\l).
\ee
One might expect that at the lowest non-trivial order the energy could
depend on the relative position of the solitons $z$, but this dependence
cancels:
\be
E_1 = \frac{4\pi}{g^2}\l^2~.
\ee
(For the unstable excited states of \cite{us}, $E_1 =
\frac{2\pi}{g^2}\l^2(4n+2)$.)  The $z$-independence at the
lowest nontrivial order implies there is a range of energies
for which the moduli space is an accurate description even at finite
$\th$. At next order w find
\be
\label{U_effect}
E_2 = -\frac{\pi\l^2}{g^2\th}\frac{V^{\prime\prime}(
\l)+V^{\prime\prime}(0)}
{V^{\prime\prime}(\l)V^{\prime\prime}(0)}
\left\{2 + \left(\frac{2z\bar{z}}{\sinh(2z\bar{z})}\right)^2
\right\}~.
\ee
The $z$-dependent part of this produces an attractive force between the
solitons. However, it is very short-range, vanishing as $e^{-4|z|^2}$.
For small $\abs{z}$, the potential between the solitons goes
smoothly to a finite constant. (One may consider solitons such that the
false vacuum is a maximum rather than a minimum: $V''(\l)<0$.  Then
the potential between the solitons could be repulsive or vanish; however,
in this case the solitons are unstable.)

\subsection{Corrections to the moduli space metric}
Using the perturbative scheme developed above,
we can compute the leading corrections to the level zero $n$ soliton
metric (\ref{nsol_metric}). To order $\th^{-1}$ we
have:
\be
g_{\a\bar\b} = \frac{1}{\l^2}{\rm Tr}
\left(\d_{\a}\Phi\db_{\b}\Phi\right)
= g^{(0)}_{\a\bar\b} +
\frac{2}{\th}g^{(1)}_{\a\bar\b} + {\cal O}(\th^{-2})~,
\ee
where
\be
g^{(1)}_{\a\bar\b} = \frac12\left({\rm Tr}
\left(\d_{\a}P_n\db_{\b}B_n\right) +{\rm Tr}
\left(\d_{\a}B_n\db_{\b}P_n\right)\right)~,
\ee
\be
B_n = \left(\frac{1}{V''(\l)}P_n + \frac{1}{V''(0)}(1-P_n)\right)
[a\, ,[\, P_n\, ,a^\dag\, ]\,]~,
\ee
and $\d_{\a} \equiv \d_{z^\a}$ and
$\db_{\b} \equiv\d_{\bar z^\b}$.
For
\be
P_n=\sum\limits_{\a ,\b =1}^{n}
\left|z_\a\right\rangle A^{-1}_{\a \b}
\left\langle z_\b\right|
\ee
we find:
\be
\label{metrcorr}
g^{(1)}_{\a\bar\b} & = & \frac{1}{V''(\l)}
\left(z_\a A^{-1}_{\a\b}\bar z_\b
- A^{-1}_{\a\b} - \sum_{\l,\k = 1}^{n}
A^{-1}_{\a\l}\bar z_\l A_{\l\k}z_\k
A^{-1}_{\k\b}\right)\times \nonumber \\
&& \hskip 15mm \times
\left(\bar z_\b A_{\b\a}z_\a
+ A_{\b\a} - \sum_{\c,\l = 1}^{n}
A_{\b\c}z_\c A^{-1}_{\c\l}\bar z_\l
A_{\l\a}\right)  \\
&& -  \;\frac{1}{V''(0)} A^{-1}_{\a\b}\left(A_{\b\a}
+ 2\bar z_\b A_{\b\a}z_\a
+ \sum_{\l,\k,\c,\l = 1}^{n}
A_{\b\l}z_\l A^{-1}_{\l\k}\bar z_\k
A_{\k\c}z_\c A^{-1}_{\c\l}\bar z_\l
A_{\l\a}\right.
\nonumber \\ &&
\left. \hskip 15mm - \;\sum_{\sigma,\k = 1}^{n}
\left(2 + z_\a\bar z_\k + z_\l\bar z_\b - z_\a \bar
z_\b\right)
A_{\b\l}z_\l A^{-1}_{\l\k}\bar z_\k
A_{\k\a}\right) \nonumber~.
\ee
To find the metric on the two-soliton relative moduli space, we set $z^1
= z$, $z^2 = -z$, and hence
\be
g_{z\bar z}~ \propto ~ (g_{1\bar1}-g_{1\bar2})~.
\ee
Explicitly,
\be
A = \left(\begin{array}{cc}
1 & {\rm e}^{-2|z|^2} \\
{\rm e}^{-2|z|^2} & 1
\end{array}\right)~,
\ee
and (\ref{metrcorr}) gives
\be
\label{g1}
g^{(1)}_{z\bar z} = \frac{\coth r^2}{V''(\l)}\left[
\left(\frac{r^2}{\sinh r^2}\right)^2-1\right] +
\frac{\coth r^2}{V''(0)}\left[\frac{4r^2}{\sinh 2r^2}
- \left(\frac{r^2}{\sinh r^2}\right)^2-1\right]~,
\ee
where $r^2 = 2z\bar z$. Note that for $V^{\prime\prime}\left(\l\right) =
V^{\prime\prime}\left(0\right)$, this correction to $f$ takes exactly
the same functional form as the original $f$.

\section{Stability analysis}
In \cite{basic} it was shown that there
exists a path in field space interpolating between
field configurations corresponding to the operators
$|n\rangle\langle n|$ and $|0\rangle\langle 0|$ and along which
the gradient energy decreases monotonically. This path is given by
$|\a\rangle\langle\a|,$ $0\leq\a\leq\frac\pi2,$ where
\be
|\a\rangle = \cos\a\,|n\rangle + \sin\a\,|0\rangle~.
\ee
Hence for finite
$\th$ the state $|n\rangle\langle n|$ decays to the state
$|0\rangle\langle 0|$.

However, for finite $\th$ the state
$|n\rangle\langle n|$ differs from the stationary point of the
full static energy functional by terms of order $\th^{-1}$,
and the energy of the true solution is smaller then the energy corresponding
to the $|n\rangle\langle n|$ state by terms of the same order.
Due to this energy difference the true level $n$-soliton cannot decay along
the $|\a\rangle\langle\a|$ path. Its instability can be, however,
still demonstrated with the help of the perturbative methods discussed
in the section \ref{Perturbation_Theory}.

We define
\be
P_\a \equiv |\a\rangle\langle\a|~,\qquad
\De_\a = [a\, ,[P_\a\, ,a^\dag\,]\,]~.
\ee
$P_\a$ satisfies the consistency condition
\be
[\, P_\a\, ,\De_\a\,] = 0
\ee
only for $\a$ an integer multiple of $\pi\over2$, but we can still
consider a path in field space given by:
\beq
\label{path}
\Phi(\a) = \l\left[P_\a + \frac{1}{\th}\left(
\frac{1}{V''(\l)}P_\a\De_\a P_\a
+ \frac{1}{V''(0)}(1-P_\a)\De_\a(1-P_\a)
\right)\right]~.
\equ
$\Phi(0)$ and $\Phi(\pi/2)$ are (up to the terms of order
$\th^{-2}$) the true level $n$- and level $0$-solitons, and the energy
of the $\Phi(\a)$ configuration is given by (for $n > 1$)
\begin{eqnarray*}
E(\a)  & = &  \frac{2\pi\th}{g^2}V(\l) +
\frac{\pi\l^2}{g^2}\left[1 + 2n\cos^2\a
-\frac{1}{\th V''(\l)}\left(1 + 2n\cos^2\a\right)^2
\right. \\
&& \left.\hskip 4cm
-\;\frac{1}{\th V''(0)}\left(1 + 2n\cos^2\a+ 2n^2\cos^4\a\right)
\right] + {\cal O}\left(\th^{-2}\right)~.
\end{eqnarray*}
When $\th V''\gg n$, the function $E(\a)$ decreases
monotonically for $0\leq\a\leq\frac{\pi}{2}$, and hence the
$\proj{n}$ soliton is unstable for $n > 1$.  For $n=1$ the
resulting formula for $E(\a)$ is slightly different, but the
conclusion is the same.

The analysis above can be extended to the case of the solutions corresponding
at $\th = \infty$ to the projection operators
\be
P_{I} = \sum_{k\in I}\proj{k}~,
\hskip 1cm I\subset I\!\!N~.
\ee
If there is a ``gap'' in the set $I$, {\it i.e.}, for
some $m < n$ we have $m\not\in I$ and $n\in I$, then using the
projectors
\be
P_{m,n}(\a) =
\left(\cos\a\, \ket{n} + \sin\a\, \ket{m}\right)
\left(\bra{n}\cos\a  + \bra{m}\sin\a\right)~,
\ee
we can construct a ``decay path'' as in (\ref{path}). This shows
that potentially stable, radially symmetric, level $n$-solitons $\phi_{(n)}$
must approach
\be
\phi^{\infty}_{(n)}= \l \sum_{k=0}^{n}\proj{k}
\ee
as $\th \to \infty$. We now consider the stability of such states.

For $\psi$ of the (general) form
$$
\psi\left(\{f_k\},U\right) = \l\sum\limits_{k=0}^\infty
f_k\ U^\dag|k\rangle\langle k|U~,
$$
where $U$ is some unitary operator, the energy functional
(\ref{energy}) can be written as
\be
\label{en_gen_1}
E[\psi]  & = &  \frac{2\pi}{g^2}\sum\limits_{k=0}^\infty
\left[\th V(\l f_k) + \frac12\l^2 f_k^2\right]
        +  \frac{2\pi\l^2}{g^2}
\sum\limits_{k=0}^\infty\sum\limits_{l=0}^\infty
lf_k^2\left|U_{k,l}\right|^2 \nonumber
\\
&&\qquad\qquad
-  \;\frac{2\pi\l^2}{g^2}
\sum\limits_{k=0}^\infty\sum\limits_{l=0}^\infty f_kf_l
\left|\sum\limits_{p=0}^\infty\sqrt{p+1}U_{k,p}U^*_{l,p+1}\right|^2~,
\ee
with $U_{p,k} \equiv \langle p|U|k\rangle$.

The radially symmetric states $\phi_{(n)}$ have the form
\be
\label{phi_n}
\phi_{(n)} = \l\sum\limits_{k=0}^\infty\ c_k\proj{k}
\ee
where, for large enough $\th$, the perturbative analysis gives
\beq
\label{pert_sol}
c_k = \left\{
\begin{array}{ll}
1 + {\cal O}\left(\th^{-2}\right) & k < n~, \\
& \\
1 - \frac{2(n+1)}{\th V''(\l)} + {\cal O}\left(\th^{-2}\right)\qquad\qquad
&
            k = n~, \\
& \\
\frac{2(n+1)}{\th V''(0)} + {\cal O}\left(\th^{-2}\right) &
            k = n+1~, \\
& \\
{\cal O}\left(\th^{-2}\right) & k > n+1~.
\end{array}
\right.
\equ
To check stability of this solution we consider
\be
\phi_{(n)} + \de\phi = \l\sum\limits_{k=0}^\infty\
(c_k+\varepsilon\de c_k)\, U^\dag
\proj{k}U~,
\ee
where $\de c_k$ are arbitrary real parameters and
$U\equiv {\rm e}^{i\varepsilon T}$ with arbitrary hermitean $T.$ Using
(\ref{en_gen_1}) we find
\be
\label{energy_pert}
&& \hskip -.5cm \frac{g^2}{2\pi}E[\phi_{(n)} + \de\phi]
=  \th\sum\limits_{k=0}^\infty V(c_k)
+ \l^2\sum\limits_{k=0}^\infty\left(\left(k+\frac12\right)c_k^2 -
\left(k+1\right)c_kc_{k+1}\right) \nonumber \\
&& + \; \frac12\varepsilon^2\l^2\sum\limits_{k=0}^\infty
\left(\th V''(\l c_k)\left(\de c_k\right)^2 +(k+1)
\left(\de c_{k+1} - \de c_k\right)^2\right) \\
&&  + \;\varepsilon^2\l^2
\sum\limits_{k=0}^\infty\sum\limits_{l=0}^\infty
\left\{(l-k)c_k^2\left|T_{k,l}\right|^2
-c_kc_l\left|\sqrt{l}T_{k,l-1} - \sqrt{k+1}T_{k+1,l}\right|^2
\right. \nonumber \\
&& \left.
+\; c_k c_{k+1}\left[(k+1)\left(\left|T_{k,l}\right|^2 +
\left|T_{k+1,l}\right|^2\right) -
\sqrt{(k+1)(l+1)}\left(T_{k,l}T_{k+1,l+1}^* + {\rm c.c.}\right)
\right]\right\} + {\cal O}\left(\varepsilon^3\right).
\nonumber
\ee
The second line of this equation
shows that as long as $V''(\l c_k) > 0$ for all $k$,
$\phi_{(n)}$ is stable against ``radial''
perturbations $c_k \to c_k + \varepsilon\de c_k$.

To check stability
against unitary rotations we use the perturbative form of
$\phi_{(n)}$ (\ref{phi_n},\ref{pert_sol}). We denote
by ${\cal E}$ the $T$-dependent part of
(\ref{energy_pert}). Terms of order $\th^0$ in ${\cal E}$
can be written in the manifestly positive semi-definite form:
\beq
\label{theta0}
{\cal E}|_{\th^0} = \sum_{k=0}^{n}\sum\limits_{l = n+2}^{\infty}
\left|\sqrt{l}T_{k,l} - \sqrt{k}T_{k-1,l-1}\right|^2~.
\equ
In general, these dominate any terms that are lower order in $\th$;
however, (\ref{theta0}) has zero-modes, so we need to consider terms
of order $\th^{-1}$:
\be
\label{theta1}
{\cal E}|_{\th^{-1}} & = & \frac{2(n+1)}{\th V''(\l)}\left[
\sum\limits_{k=0}^n\left|\sqrt{n+1}T_{k,n+1}-\sqrt{k}T_{k-1,n}\right|^2
- \sum\limits_{l=0}^\infty (l-n)|T_{l,n}|^2\right] \nonumber\\
&& + \; \frac{2(n+1)}{\th V''(0)}\left[
\sum\limits_{l=n+1}^\infty\left|\sqrt{n+1}T_{n,l}-\sqrt{l}T_{n+1,l+1}\right|^2
- \sum\limits_{l=0}^\infty (l-n-1)|T_{l,n+1 }|^2\right] \nonumber \\
&& - \;\frac{2(n+1)}{\th V''(\l)}\sum\limits_{l=n+2}^\infty
\left|\sqrt{l}T_{n,l}- \sqrt{n}T_{n-1,l-1}\right|^2  \\
&& - \; \frac{2(n+1)}{\th V''(0)}\sum\limits_{k=0}^n
\left|\sqrt{n+2}T_{k,n+2}-\sqrt{k}T_{k-1,n+1}\right|^2~. \nonumber
\ee
Some of these are positive as they stand, and some are obviously
dominated by the ${\cal O}(\th^0)$ terms, {\it e.g.}, the negative sum
proportional to $\frac{1}{V''(\l)}$ together with the terms in
(\ref{theta0}) with $k=n$ can be written in the form
$$
\left(1- \frac{2(n+1)}{\th V''(\l)}\right)\sum\limits_{l=n+2}^n
\left|\sqrt{l}T_{n,l}- \sqrt{n}T_{n-1,l-1}\right|^2~,
$$
which is positive for $\frac{2(n+1)}{\th V''(\l)} < 1$.
Using the identity
$$
\left|\sqrt{n+1}T_{n,n+1}-\sqrt{n}T_{n-1,n}\right|^2
+\left|T_{n-1,n}\right|^2 -\left|T_{n,n+1} \right|^2
=
\left|\sqrt{n}T_{n,n+1}-\sqrt{n+1}T_{n-1,n}\right|^2
$$
we can rewrite the remaining, proportional to $\frac{1}{V''(\l)}$ terms in
(\ref{theta1})  as
\be
\label{thetaneg}
&& \frac{2(n+1)}{\th V''(\l)}
\left\{\left|\sqrt{n}T_{n,n+1}-\sqrt{n+1}T_{n-1,n}\right|^2 +
\sum\limits_{k=1}^{n-1}\left|\sqrt{n+1}T_{k,n+1}-\sqrt{k}T_{k-1,n}\right|^2
\right. \nonumber \\
&& \left. \hskip 2cm +\; (n+1)|T_{0,n+1}|^2 +
\sum\limits_{k=0}^{n-2}(n-k)|T_{k,n}|^2 -
\sum\limits_{l=n+2}^\infty (l-n)|T_{n,l}|^2
\right\}.
\ee
These terms are potentially dangerous only
along the zero mode direction of ${\cal E}|_{\th^{0}}$ with
some nonvanishing $T_{n,l}, l\geq n+2$.  Such modes obey the condition
\beq
\label{cond_zero}
        \sum_{k=0}^{n}\sum\limits_{l = n+2}^{\infty}
\left|\sqrt{l}T_{k,l} - \sqrt{k}T_{k-1,l-1}\right|^2 = 0~,
\equ
which gives $\left|T_{0,l}\right|=0$ for $l \geq n+2$ and the recursion
relation
\beq
\label{recursion}
T_{k,l} = \sqrt{\frac{k}{l}}T_{k-1,l-1}~;
\equ
this implies
$$
T_{n,l} = 0 \hskip 5mm {\rm for} \hskip 5mm l > 2n+1.
$$
For $l=2n+1$ (\ref{recursion}) gives
$$
(n+1)\left|T_{n,2n+1}\right|^2 = \frac{\left((n+1)!\right)^2}{(2n+1)!}
\left|T_{0,n+1}\right|^2.
$$
while for $l = n+p$ with $2\leq p \leq n$ we get
$$
\left|T_{n,n+p}\right|^2 = \frac{(n!)^2}{(n-p)!(n+p)!}\left|T_{n-p,n}\right|^2
$$
and the last three terms in (\ref{thetaneg}) give
\begin{eqnarray*}
&&
(n+1)|T_{0,n+1}|^2 +
\sum\limits_{k=0}^{n-2}(n-k)|T_{k,n}|^2 -
\sum\limits_{l=n+2}^\infty (l-n)|T_{n,l}|^2 \\
&&  = \;(n+1)\left(|T_{0,n+1}|^2 - \left|T_{n,2n+1}\right|^2\right)
+ \sum\limits_{p=2}^n p
\left(\left|T_{n-p,n}\right|^2- \left|T_{n,n+p}\right|^2\right) \\
&& = \;\left( n+1 - \frac{\left((n+1)!\right)^2}{(2n+1)!}\right)
\left|T_{0,n+1}\right|^2 + \sum\limits_{p=2}^n p
\left(1-\frac{(n!)^2}{(n-p)!(n+p)!}\right)
\left|T_{n-p,n}\right|^2 > 0~.
\end{eqnarray*}
Similar analysis applies to the terms proportional to $\frac{1}{V''(0)}.$

The remaining zero modes of (\ref{energy_pert}) (apart from the exact,
translational zero mode which  -- for $\th \to \infty$ 
-- is just $T_{n,n+1}$) 
are given by $T_{k,l}$ with $k > n+1$ or
with $l < n$ (thanks to the hermiticity of $T$ we can always choose
$k$ to be smaller than $l$). The first set of these
zero modes (with $k > n$) is irrelevant -- they correspond to
the unitary transformations that do not change the state to order $\th^{-1}$.

The other set of zero modes (with $k < l < n$) corresponds to
rotations that act on the nonzero $c_k$'s and do change the soliton. To check
if they can destabilize the soliton we would have to extend the perturbative
analysis to higher orders in $\theta^{-1}$. Fortunately, they are
absent for the most interesting case, that of two solitons (i.e. for $n=1$).

\section{The geodesic equation at finite $\th$}
We now study classical scattering in the presence of a
generated potential $U$.
\subsection{Integrating the classical equations of motion}
For concreteness and simplicity we consider the two-soliton case
(\ref{U_effect}):
\be
U(r) = -\left(\frac{1}{2\th}\right)\frac{V''(
\l)+V''(0)}
{V^{\prime\prime}(\l)V^{\prime\prime}(0)}
\left(\frac{r^2}{\sinh(r^2)}\right)^2~.
\ee
The effective action for the two-soliton system is
\be
\frac{2\pi\l^2}{g^2} \int dt \left( \frac{\th}{2}\left(g_{rr}\dot{r}^2 +
g_{\vartheta\vartheta}\dot{\vartheta}^2\right) -U(r) \right)~,
\ee
where the metric $g$ is given by $ds^2 = f(r)(dr^2 + r^2 d\vartheta^2)$
and $f$ is as in \cite{us}
\be
\label{f}
f(r) = \coth(r^2) - \frac{r^2}{\sinh^2(r^2)}~.
\ee
In principle, we should consider the corrections (\ref{g1}) to the moduli
space metric; as discussed below, for large $\th$, these can be ignored. In
contrast, the potential $U$ is important even though it is also suppressed
at large $\th$.

Varying the action leads to the equations of motion
\be
\ddot{r} + \Gamma^{r}_{rr}\dot{r}^2 + \Gamma^{r}_{\vartheta\vartheta}
\dot{\vartheta}^2 +\frac{g^{rr}}{\th}\frac{dU}{dr} = 0 ~,\\
\nonumber\\
\ddot{\vartheta} + 2\Gamma^{\vartheta}_{r\vartheta}\dot{r}\dot{\vartheta}
= 0 ~.
\ee
The second equation is the same as for the case without $U$ and the
solution is \cite{us}
\be
\dot{\vartheta} = \frac{l}{r^2 f(r)}
\ee
where $l$ is an integration constant corresponding to the angular
momentum. The first equation can then be written as
\be
\ddot{r} + \frac{1}{2}\frac{d}{dr}\ln(f) \dot{r}^2 =
\frac{1}{2f}\frac{d}{dr}(r^2 f)\left(\frac{l}{r^2 f}\right)^2
- \frac{1}{\th f}\frac{dU}{dr}~.
\ee
An integrating factor for this equation is $\dot{r}f$ from which one
finds the solution
\be
f\dot{r}^2 = 2\frac{E - U}{\th} - \frac{l^2}{r^2f}~,
\ee
where $E$ is an integration constant with the interpretation of the
total energy of the system.  In the same way as in the case without a
potential \cite{us}, this leads to scattering trajectories found from the
integral
\ber
\vartheta(r) = - \int_{\infty}^{r}
\frac{ds}{s\sqrt{\left(\frac{s}{b}\right)^2 f(s)(1-\frac{U}{E})-1}}
\eer{modgeo}
here $b^2 = \frac{l^2 \th}{2E}$ is the impact parameter as in
\cite{us} (the $\th$ dependence arises because we have rescaled the
coordinates). The finite $\theta$ correction to the geodesic
scattering picture can therefore be found by using a corrected function
\be
\tilde{f}(r) = f(r)\left(1-\frac{U(r)}{E}\right)~.
\ee
Since $U$ is attractive (negative) there are no extra
divergencies in the effective $\tilde{f}$ as compared to $f$. If $U$ had
been repulsive, but of the same functional form\footnote{This could happen
for a potential $V$ where the false vacuum corresponds to a maximum;
however, such solitons are unstable.}, it would have made the effective
$\tilde{f}$ more repulsive.

We can make some estimates of the validity of our approximations by
restoring the dimensions of the coordinates: $r \to \frac{r}{\sqrt{\th}}$.
Since $E = \frac{\l^2}{g^2}(\frac{f}{2}v^2 + U)$,
which corresponds to a particle of effective mass $f\frac{\l^2}{g^2}$
moving in a potential $U$, we can find a range of velocities where the moduli
space approximation should be good. For the correction to the classical
result in (\ref{modgeo}) to be small we need $\frac{U}{E} \ll 1$ leading to
\be
v^2 \gg \frac1{2 \th}
\frac{V^{\prime\prime}(\l)+V^{\prime\prime}(0)}
{V^{\prime\prime}(\l)V^{\prime\prime}(0)}.
\ee
However, for the adiabatic approximation to be valid, the momentum
transfer must remain sufficiently small so that fluctuations out of the
moduli space are suppressed. In our case there are several possiblities since
potentials for different fluctuations appear at different orders in
perturbation theory. Even if we do not have the exact potentials, we can
estimate their strength from the general behavior of perturbation theory.
Most fluctuations have potentials already at $\th =\infty$. They are not
excited as long as
\be
v^2 \ll \frac{{\th} V(\l)}{\l^2}.
\ee
Other fluctuations get a potential only at first order in
perturbation theory. They are not excited as long as
\be
v^2 \ll 1~;
\ee
this simply means that the motion remains nonrelativistic. For the
two-soliton case, we have checked that the fluctuations with lower energies
correspond to motions of the solitons, which we do not want to restrict.
Higher soliton scattering requires a higher-order analysis.

\subsection{Trajectories}

It is interesting to investigate some explicit cases for the
scattering trajectories of the previous section. We have prepared
movie clips in MPEG format\footnote{If the reader's viewer does not
support hypertex, the three movies can be found at {\tt
http://www.physto.se/\~{}unge/traj1.mpg},
{\tt http://www.physto.se/\~{}unge/traj2.mpg}, and
{\tt http://www.physto.se/\~{}unge/traj3.mpg}.}.
The$\!\!$
{\catcode`\~=11
}
first movie

$\!\!$shows the behavior for
large values of the impact parameter. The solitons just pass each
other with no scattering taking place. In the$\!\!$
{\catcode`\~=11
}
second movie

$\!\!$the right angle scattering for small impact parameter $b$ is
shown\footnote{The second movie has time slowed down by a factor of 1000; to
keep the file a manageable size, a smaller spatial region is shown.}. Notice
that this qualitative behavior is true irrespective of the value of the total
energy $E$ since it only depends on the value of the function
$\tilde{f}$ at large or small $r$. However, in the presence of the attractive
potential
$U(r)$ and for small enough energy ($\frac{U_0}{E} > 3.86$) we find new
qualitative behavior shown in the$\!\!$ {\catcode`\~=11
}
third movie
$\!\!$.
We get a metastable orbit where the solitons circle
around each other for some time before they scatter to
infinity. These results are summarized in the following
picture where the exit angle is plotted as a function of the impact
parameter in the case where $\frac{U_0}{E} = 5$.

\vskip 5mm
\centerline{\epsfxsize=9cm\epsfbox{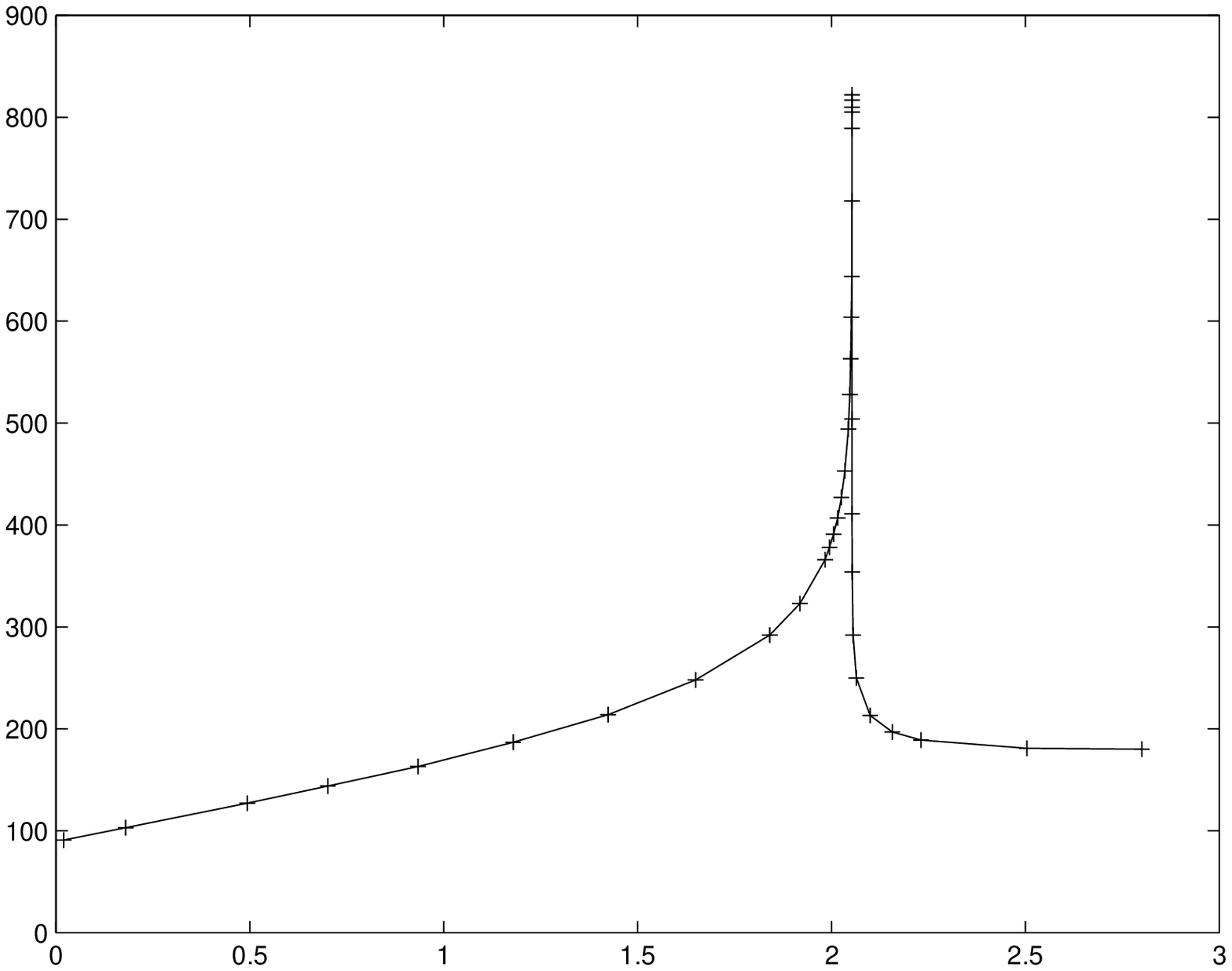}} \nopagebreak
\vskip 2mm \nopagebreak
\noindent
{\bf Figure 1:} The exit angle in degrees as a function of the impact
parameter for $\frac{U_0}{E} = 5$.

\vskip 5mm

One may also ask what happens when we include the corrections to the
moduli space metric (\ref{g1}). When $V''(\l) = V''(0)$, $f$ is rescaled
by $(1+\frac{2\l}{\th})$, which can be absorbed by a redefinition of $b$.
However, when $V''(\l)$ and $V''(0)$ are not equal we find two different
behaviors: If $V''(\l) > V''(0)$, the trapped orbit effect above is
suppressed, that is, it appears for smaller $E$, or larger $\frac{U_0}{E}$,
than before. On the other hand, if $V''(0) > V''(\l)$ we get an enhancement of
the effect. In fact, trapped orbits appear even for $U_0 = 0$ if $V''(\l)$ is
small enough!

\section{Quantization}

In this section we discuss the quantization of the effective hamiltonian
that describes the motion of solitons. We focus on the two-soliton case.
The Schr\"{o}dinger equation for this problem can be written
as\footnote{There is really a factor of $\frac{1}{2\sqrt{\th}}$ in front of
the $\nabla^2$ operator but we can soak it up in a redefinition of $U$ and
$E$ so that they become dimensionless.}
\be
\left(-\nabla^2 + U \right)\psi = E \psi~,
\ee
where
\be
\nabla^2 = \frac{1}{fr}\d_{r}\left(r\d_{r}\right) +
\frac{1}{fr^2} \d_{\vartheta}^2~,
\ee
and $f(r)$ is the metric (\ref{f}).
In the absence of a potential, this Hamiltonian operator is
positive\footnote{We thank Matthew Headrick of reminding us of this fact and
pointing out that by ignoring it in a previous version of this paper, we had
made a serious error.}; consequently, it cannot have any bound states.  For any
attractive potential, no matter how small, an $s$-wave bound state appears.
Thus the potential that we found perturbatively (\ref{U_effect}) induces such
a bound state for any finite value of $\th$. The potential is rotationally
symmetric and hence for
$\psi =
\chi(r)e^{il\vartheta}$ the equation reduces to:
\be
\left( -\frac{1}{fr}\d_{r}\left(r\d_{r}\right)
        + \frac{l^2}{fr^2} +U\right)\chi = E\chi~.
\ee
We may estimate the energy of the bound state as follows: for sufficiently
small $r$, $f\approx \frac23 r^2$ and $U\approx U_0<0$. This admits a solution
with energy $E$ of the form of a Bessel function $J_0(kr^2)$ for $k^2
= (U_0-E)/6$. For large r, $f\approx 1$ and $U\approx 0$; this admits a
solution of the form $A\, K_0(\sqrt{-E}\,r)$, where $A$ is some normalization
constant and $K_0$ is a Bessel function of the second kind. Matching these at
some intermediate value $r_m$ where $U(r_m)\to E$ gives
\be
\frac{2kr_mJ_1(kr_m^2)}{J_0(kr^2_m)} ~=~
\frac{\sqrt{-E}\,K_1(\sqrt{-E}\,r_m)}{K_0(\sqrt{-E}\,r_m)}~;
\ee
for small enough $|U_0|,|E|$, we find
\be
E ~ \approx ~ -\,\frac1{r_m}\,e^{-2/(U_0r_m^4)}~.
\ee
Of course, the details of $f(r)$ and $U(r)$ correct the solution and the
energy, but they cannot change the qualitative behavior.

Because of this $s$-wave bound state,
for sufficiently small soliton energy the cross section becomes very
large; this ruins the moduli space picture. The conclusion is therefore the
same as in the previous section: for the moduli space picture to be a good
approximation we need energies in an intermediate range, not big enough to ruin
the adiabatic approximation but not so small as to see the bound
state.

The question of how the classical metastable states found in the previous
section appear in the quantum treatment is left to future work.

\acknowledgments{We are happy to thank Alfred Goldhaber and Erick Weinberg
for discussions. We thank Matthew Headrick for pointing out that our
original analysis of the Schr\"odinger equation was incorrect, and Dennis
Sullivan for helping us understand the correct picture. The work of LH was
supported by a Foundation for Polish Science fellowship. The work of UL was
supported in part by NFR grant 5102-20005711 and by EU Contract
HPRN-C7-2000-0122. The work of MR was supported in part by NS grant
PHY9722101. The work of RvU was supported by the Czech Ministry of Education
under Contract no. 143100006.}

\vskip .4cm

\noindent{\bf Note added:} After completing our work, \cite{GHS} appeared.
This work has significant overlap with ours. The authors observe that
the solitons naturally obey Bose-Einstein statistics, and that imposing this on
the solitons replaces the moduli space with its quotient by the symmetric
group; as pointed out in \cite{us} (where, however, we did {\it not} impose
Bose-Einstein statistics), this quotient smoothes the singularities discussed
in section 2. In the quantum analysis of section 6, Bose-Einstein statistics
require us to consider only even angular momentum $l$. Finally, we want to thank
the referee for making us aware of \cite{MM}, where noncommutative solitons are 
considered from a somewhat different perspective.

\end{document}